\documentclass{amiacri}

\usepackage{url}
\usepackage{algpseudocode}
\usepackage{algorithm}
\usepackage{graphicx}
\usepackage[labelformat=simple]{subcaption}
\usepackage{amsmath}
\usepackage{multirow}
\usepackage[usenames,dvipsnames,table]{xcolor}
\usepackage{colortbl}

\graphicspath{{./figures/}} 

\newcommand{\trc}[1]{\multirow{2}{*}{#1}} 
\definecolor{lg}{gray}{0.9}
\definecolor{up}{RGB}{255,77,77}
\definecolor{down}{RGB}{77,255,77}

\begin{document}

\title{A Simple Text Mining Approach for Ranking Pairwise Associations in Biomedical Applications}

\author{
Finn Kuusisto, PhD$^1$,
John Steill, MS$^1$,
Zhaobin Kuang, MS$^2$,\\
James Thomson, VMD, PhD$^{1,2}$,
David Page, PhD$^2$,
Ron Stewart, PhD$^1$
}

\institutes{
$^1$Morgridge Institute for Research, Madison, USA
$^2$University of Wisconsin, Madison, USA
}

\maketitle

\section*{Abstract} \vspace{-1em}
{\itshape
We present a simple text mining method that is easy to implement, requires minimal data collection and preparation, and is easy to use for proposing ranked associations between a list of target terms and a key phrase.
We call this method KinderMiner, and apply it to two biomedical applications.
The first application is to identify relevant transcription factors for cell reprogramming, and the second is to identify potential drugs for investigation in drug repositioning.
We compare the results from our algorithm to existing data and state-of-the-art algorithms, demonstrating compelling results for both application areas.
While we apply the algorithm here for biomedical applications, we argue that the method is generalizable to any available corpus of sufficient size.
}

\section*{Introduction} \vspace{-1em}
Many scientific discoveries are often subject to lengthy processes of trial and error before important and meaningful results are found.
For example:
\begin{enumerate}
\item In biology, determining a set of defined transcription factors for differentiating or reprogramming cell types requires trying numerous combinations from lists of factors.
The combinatorial growth of the search space quickly leads to intractability.
\item In medicine, discovering off-label uses of approved drugs can take years of collecting observational data and running post-approval trials.  Once again, the search becomes time-consuming due to the enormous number of pairs of drugs and effects.
\item Similarly in medicine, detecting adverse drug events can require extensive observational data to detect potential correlations between drugs and events.
\end{enumerate}

Because the search spaces are so large, proper prioritization of research directions in these cases is essential to reaching novel discoveries quickly, but this requires both extensive breadth and depth of knowledge within the domain.
Furthermore, due to exponential growth in scientific literature,\cite{pautasso2012publication,bornmann2015growth} it is becoming continually more challenging to keep up with current knowledge in any particular domain.
We present a general text mining approach to address this prioritization problem by ranking a list of target terms (e.g. transcription factors or drugs) by their association with a key phrase (e.g. ``embryonic stem cell'' or ``hypoglycemia'').
This list provides researchers with a starting point for entering the literature domain and prioritizing potential research directions, thereby accelerating the discovery process.
Our method is easy to implement, requires minimal data collection and preparation, and is easy to use.

To produce our ranked list of target terms associated with a key phrase, we leverage the vast collective knowledge available within the published scientific and medical literature.
We use simple keyword matching and document counting to automatically identify significant correlations and rank them by their co-occurrence proportion.
Owing to its simplicity, we call our method KinderMiner.

While we can imagine several applications of our approach, we focus our attention on the two former examples given above: determining important transcription factors for cell reprogramming and discovering off-label uses of approved drugs.
To assess our approach, we compare rankings produced by our approach with three cell reprogramming tasks that have experimentally proven sets of defined factors from landmark publications.
For fairness, we censor the literature in our experiments to publications from roughly two years prior to the relevant landmark publications.
We also apply our approach to the task of discovering drugs that may be repurposed for reducing blood glucose.
In both cases, we show that our method is able to reproduce sufficient sets of defined factors and many relevant drugs within the top hits, suggesting that our method will likely be useful in accelerating the discovery process.

\section*{The KinderMiner Algorithm} \vspace{-1em}
Algorithm \ref{alg:alg} breaks KinderMiner down step-by-step.
At a high level, KinderMiner ranks a list of target terms by their association with a specified key phrase.
It does this via keyword matching and document counting within a specified, relevant, searchable text corpus.

\begin{algorithm}[!ht]
  \centering
  \caption{The KinderMiner algorithm.}
  \label{alg:alg}
  \begin{algorithmic}
    \State \textbf{Input:} $\mathit{Corpus}, \mathit{TargetTerms}, \mathit{KeyPhrase}, \mathit{PThreshold}$
    \State $\mathit{topTerms} = \{\}$
    \State $\mathit{articleTotal} = \mathit{ArticleCount(Corpus)}$
    \State $\mathit{kpTotal} = \mathit{ArticleCountWith(Corpus, KeyPhrase)}$
    \For {$\mathit{term} \in \mathit{TargetTerms}$}
      \State $\mathit{targKP} = \mathit{ArticleCountWithBoth(Corpus, term, KeyPhrase)}$
      \State $\mathit{targNoKP} = \mathit{ArticleCountWith(Corpus, term)} - targKP$
      \State $\mathit{noTargKP} = \mathit{kpTotal} - targKP$
      \State $\mathit{noTargNoKP} = \mathit{articleTotal} - targKP - targNoKP - noTargKP$
      \State $p = \mathit{OneSidedFisherExact(targKP, noTargKP, targNoKP, noTargNoKP)}$
      \If {$p < \mathit{PThreshold}$}
        \State $\mathit{topTerms} = \mathit{topTerms} \cup \mathit{term}$
      \EndIf
    \EndFor
    \State $\mathit{sortedTerms} = \mathit{SortByKeyPhraseAndTermRatio}(\mathit{topTerms})$
    \State \Return $\mathit{sortedTerms}$
  \end{algorithmic}
\end{algorithm}

First, KinderMiner requires a large corpus of documents for querying.
While we focus on corpora of scientific literature, the corpus could also be a collection of plain text patient records taken from an electronic health record, a twitter feed, blog posts, or any other large indexed collection of plain text documents.
The corpus must be queryable for document counts with exact matching of words and phrases.
For evaluation purposes, it is also useful if the document queries can be date censored, reducing counts of documents to only those that have been published within a specified date range.
This is not required, however.

\begin{figure}[!ht]
  \centering
  \includegraphics[width=0.6\textwidth]{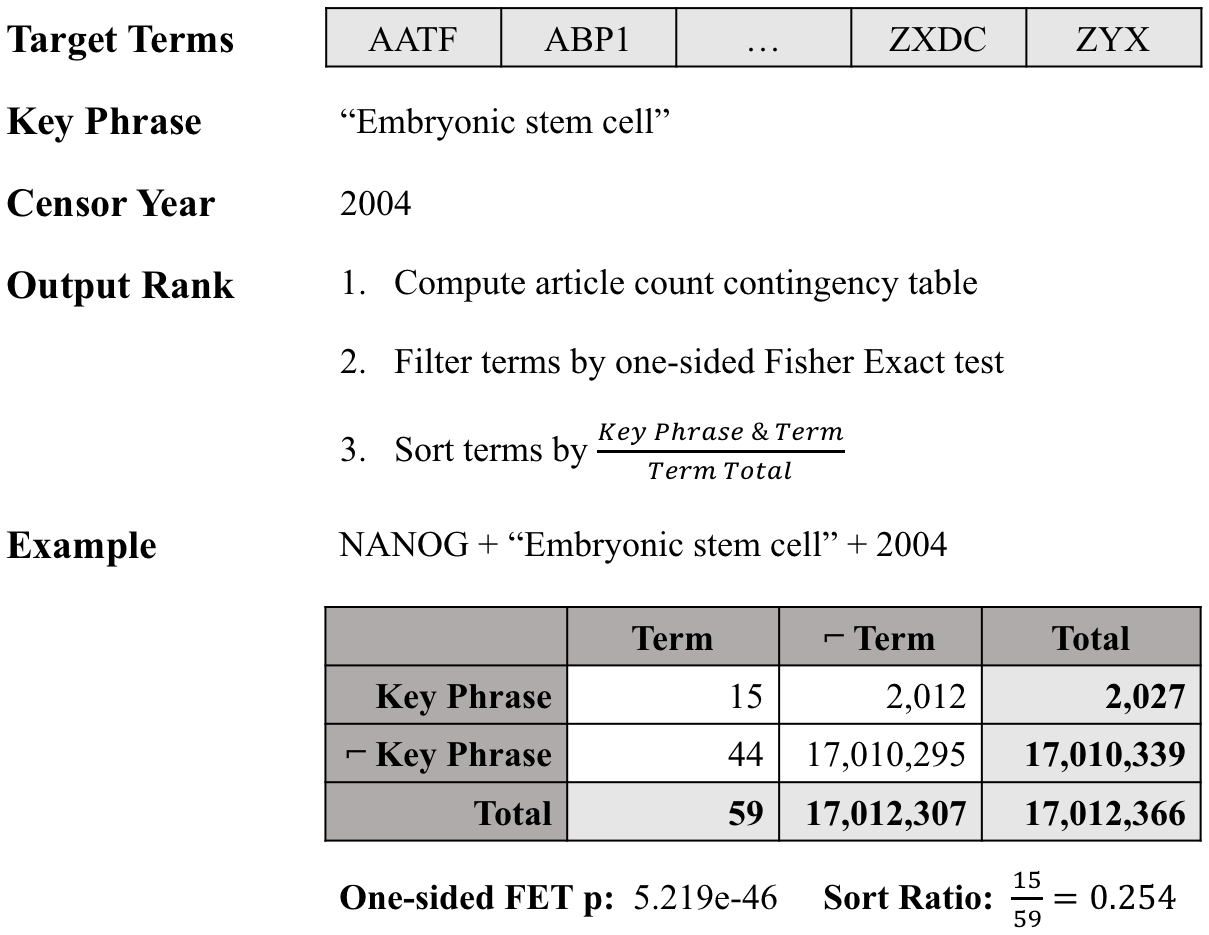}
  \caption{Visual example of KinderMiner, with contingency table and associated Fisher's Exact Test (FET) analysis of the key phrase ``embryonic stem cell'' and the target term ``NANOG.'' Target terms are filtered by significance of co-occurrence with the key phrase and then sorted by the co-occurrence ratio.}
  \label{fig:ips_example}
\end{figure}

Second, the user must specify a list of target terms to be ranked by their association with a specified key phrase.
For example, for one of our cell reprogramming applications, we specify a list of transcription factors and rank them by their association with the key phrase ``embryonic stem cell.''
The goal of this query is to identify the factors necessary for inducing an embryonic stem cell-like state.
See Figure \ref{fig:ips_example} for a more visual representation of this set of queries.

Next, for each target term, KinderMiner queries the corpus for documents that contain both, either, and neither the target term and the key phrase, producing a contingency table of document counts.
KinderMiner then performs a one-sided Fisher's exact test on the resultant contingency table, and filters out target term, key phrase pairs that do not meet a prespecified significance level.
KinderMiner uses the one-sided Fisher's exact test to assess significance only in the direction that there are more articles that contain both key phrase and target.

Finally, the selected target terms are ranked by the ratio of documents containing both the target term and the key phrase, over the total of those containing the key phrase; that is, they are ranked by the proportion of documents containing the target term that also contain the key phrase.

A great deal of work has been devoted to mining the biomedical literature.
Our simple approach is related to prior work on co-occurrence statistics and relationship extraction\cite{cohen2005survey,zweigenbaum2007frontiers} which often constrains search to particular types of relationships or relies on more sophisticated techniques such as part-of-speech tagging and named entity recognition.
KinderMiner simply constrains the search space by relying on exact text matches to an input key phrase and target terms.
Of course, KinderMiner could almost certainly benefit from NLP techniques such as text normalization and named entity recognition.
Nevertheless, our goal with this work is to address whole literature information extraction using the simplest approach we can imagine to rank potential associations, using readily available tools and sources of data, and requiring little to no data annotation or processing.
Despite its lack of sophistication, we find that our approach performs well when presented with a large corpus.

In the next two sections, we motivate two different applications, cell reprogramming and drug repositioning respectively, and evaluate the KinderMiner algorithm in the context of these applications.
We selected these particular applications not only for their significance to science and medicine, but also because of the availability of reasonable ground truth against which we can compare KinderMiner's findings.

\section*{Cell Reprogramming Applications} \vspace{-1em}
An increasingly common task in modern biology is the process of taking cells of one type and reprogramming them to exhibit the characteristics of another cell type.
Reprogramming in this case often involves introducing a set of transcription factors that put the source cells on track to behave like a different target cell type.
A particularly important example of reprogramming is that of somatic cells to an induced pluripotent stem (iPS) cell as iPS cells behave like embryonic stem cells, wherein they have the potential to differentiate into nearly all fetal or adult cell types.\cite{cellpotency2009,ips2006yamanaka,ips2007yamanaka,ips2007thomson}
Reprogramming can also be accomplished through transdifferentiation, which is when one somatic cell type is directly converted into another somatic cell type.\cite{transdifferentiation2009}
Reprogramming is important because researchers often need particular cell types to create models, study the effects of disease, develop therapies, or perform basic science, but primary cells of certain types are not always available in abundant quantities, if at all.

Altering the expression of transcription factors is also useful in the maturation of cells.
For instance, methods exist for differentiating and culturing immature hepatocytes, the main cells of the liver responsible for metabolism of drugs and toxins, but these cells are difficult to mature.
Immature hepatocytes cannot serve as reasonable surrogates for hepatocyte function, drug toxicity, or metabolism.
Recent publications\cite{hepatocyte2011,hepatocyte2013} describe methods for partial maturation of hepatocytes using transcription factors.
For similar reasons, having methods for differentiating cardiomyocytes, muscle cells of the heart, is useful, and transcription factor sets for differentiating cells into cardiomyocytes have recently been described.\cite{cardiomyocyte2010,cardiomyocyte2013}

Determining a set of important transcription factors for converting one cell type into another is, however, a challenging task that involves a great deal of domain expertise as well as trial and error.
There are roughly 2,000 transcription factors to choose from,\cite{tfcensus2009} and researchers must rely on their reading of the literature and intuition to decide which combinations to try and in what order.
This search is time consuming, and we propose that our algorithm can assist researchers by accelerating the trial and error process.
Instead of trying combinations from the entire list of transcription factors based on intuition, researchers can prioritize their experiments by exploring a much smaller number of possible combinations from only the top ranked factors provided by our algorithm.

To demonstrate our algorithm in this domain, we refer to three well-established sets of factors for reprogramming.
The first is for creating induced pluripotent stem cells (iPS cells), the second is for creating cardiomyocytes, and the third is for the maturation of hepatocytes.
We use our algorithm to mine scientific and medical literature and rank a list of transcription factors by correlation with the key phrases ``embryonic stem cell,'' ``cardiomyocyte,'' and ``hepatocyte.''
We then compare the top hits in each list with the experimentally determined factors known to produce cells representative of these cell states.
For fairness, we censor the literature available to our algorithm by roughly two years in advance of the earliest publications that demonstrate these conversions.

\section*{Drug Repurposing Application} \vspace{-1em}
Despite increases in R\&D spending, the biopharmaceutical industry has struggled to improve cost and throughput of de novo drug discovery.\cite{drugrepositioning2004}
Due to advances in key technologies and the increasing availability of data, drug repositioning, the detection of new uses for existing drugs, has become more feasible.\cite{cdrsurvey2016}
Furthermore, repositioned drugs do not require a costly development process and can reach clinical trials much faster than traditionally developed drugs.
These advantages have led repositioned drugs to constitute approximately 30\% of drugs and vaccines newly approved by the US Food and Drug Administration\cite{cdrsurvey2014}.

There have been several computational drug repositioning (CDR) approaches proposed.
Computational methods often rely on heterogeneous data sources containing genetic and phenotypic information, drug molecular structure, electronic health records, or plain-text literature as we do here.\cite{cdrsurvey2016,cdr2016charles,cdrliteraturemining2011}
We propose that our algorithm is a useful addition to the CDR toolbox, despite being far simpler than other methods.

To demonstrate our algorithm in this domain, we focus on the task of identifying drugs that may reduce blood glucose.
We use our algorithm to mine the literature and rank a list of drugs and devices by correlation with the key phrase ``hypoglycemia'' (i.e. low blood sugar).
We manually assess how well our method is able to identify drugs and devices that are specifically used to treat diabetes in the top hits, and then assess the potential of those top hits that are not specifically for treatment of diabetes.
We do not censor the date for this task.

\section*{Materials and Methods} \vspace{-1em}
For our experiments, we used the Europe PMC (EPMC) corpus.\cite{epmc2014}
We implemented our queries with EPMC's RESTful API, using the \emph{profile} search module with counts coming taken from the \emph{ALL} publication type.
We form our queries using quoted, exact matches for both the target terms and key phrases, and we use the FIRST\_PDATE parameter to censor publication year from 1900 through the specified year.
For example, a query for co-occurrence of the term NANOG and key phrase ``embryonic stem cell,'' censored to the end of 2004, would appear as follows:

\begin{center}
\texttt{``NANOG'' AND ``embryonic stem cell'' AND (FIRST\_PDATE:[1900-01-01 TO 2004-12-31])}
\end{center}

At time of writing, the EPMC corpus contains a total of approximately 27.5 million publications.
Approximately 20 million of the articles were published during or before 2008 and 17 million were published during or before 2004.

For our cell reprogramming applications, we query our lab's list of 2,243 transcription factors against the key phrases ``embryonic stem cell,'' ``cardiomyocyte,'' and ``hepatocyte.''
We use a one-sided FET p-value threshold of $1\times10^{-5}$.
We collect the top 20 transcription factors from each of these queries and use two standards for comparison.
First, we search our top factors for factors from landmark publications that have previously been shown experimentally to reprogram somatic cells to iPS cells, cardiomyocytes, and to partially mature hepatocytes.
Specifically, the relevant factors we consider for iPS cells are MYC, KLF4, LIN28, NANOG, POU5F1, and SOX2.\cite{ips2006yamanaka,ips2007thomson,ips2007yamanaka}
The relevant factors we consider for cardiomyocytes are GATA4, HAND2, MEF2C, NKX2-5, and TBX5.\cite{cardiomyocyte2010,cardiomyocyte2013}
The relevant factors for hepatocyte maturation are GATA4, HNF1A, FOXA3, FOXA2, HNF4A, CEBPB, and MYC.\cite{hepatocyte2011,hepatocyte2013}

Second, we identify our top selected transcription factors that are also indicated as being relevant by the Mogrify algorithm, a state-of-the-art algorithm to predict transcription factors for reprogramming between several cell types.\cite{mogrify2016}
Mogrify starts from gene expression data to score differentially expressed genes between cell types of interest and background expression levels.
It then combines these differential expression scores with regulatory network information to rank transcription factors in each cell type by regulatory influence.
Finally, Mogrify selects optimal sets of transcription factors with the greatest regulatory influence over differentially expressed genes in a given target cell type in comparison to a given starting cell type.
Importantly, Mogrify requires a large amount of processed data that may not be readily available and would be costly and time prohibitive to collect.

For the Mogrify comparison, we collect the complete lists of predicted transcription factors from \url{http://www.mogrify.net}.
For the iPS cell comparison, we use the conversion between \emph{dermal fibroblast} and \emph{H9 embryonic stem cells}.
For the cardiomyocyte comparison, we use the conversion between \emph{dermal fibroblast} and \emph{heart - adult}.
For the hepatocyte comparison, we use the conversion between \emph{dermal fibroblast} and \emph{liver - adult}.

For the iPS cell queries, we censor the publication date range through the end of the year 2004.
This time frame roughly corresponds to two years prior to the first publications on direct reprogramming in mouse cells.\cite{ips2006yamanaka}
For the cardiomyocyte queries, we censor the publication date range through the end of the year 2008, which also corresponds to two years prior to the first major publications on cardiomyoctye reprogramming in mice.\cite{cardiomyocyte2010}
We censor to the year 2009 for the hepatocyte applications as it corresponds to roughly two years prior to the first major publication on induction of functional hepatocytes from mouse fibroblasts.\cite{hepatocyte2011}

To evaluate our algorithm on the drug repositioning application, we query the same list of 2,609 drugs and devices used by Kuang et al.\cite{cdr2016charles} against the key phrase ``hypoglycemia'' (low blood glucose).
Again, we use a p-value threshold of $1\times10^{-5}$.
Note that we use an exact match of drug names in this case (e.g. \emph{Glucotrol} and \emph{Glucotrol XL} are treated as different) even though there may be multiple names for the same drug.
To evaluate our method, we first manually annotate the top 50 hits as either advertised specifically to treat diabetes or not.
We then compare those that were not identified as diabetes drugs to a curated list of drugs\cite{diabetesincontrol} that are known to cause hypoglycemia, hyperglycemia, or both, reporting those correctly and incorrectly identified as reducing blood glucose.
Finally, we mark the top hits that also match hits in the full list of drugs and devices predicted to reduce blood glucose by the state-of-the-art approach proposed by Kuang et al. using electronic health records.\cite{cdr2016charles}
Kuang et al. extend the self-controlled case series model\cite{farrington1995relative} to handle continuous numeric responses.
The self-controlled case series, which has been widely used for detecting adverse drug events, divides patient time-course data into control and risk periods corresponding to periods before and after exposure to a drug.
Patients thus serve as their own control cases and relative incidence of adverse events can be measured in the control and risk periods.
Importantly, this approach requires a large amount of time-course electronic health record data, which is difficult to acquire.

\section*{Results} \vspace{-1em}

\begin{table}[!ht]
\centering
\caption{Top 20 hits for each of our cell reprogramming queries. Hits that match the landmark papers are highlighted in gray, and hits that match transcription factors predicted by Mogrify are marked with *.}
\vspace{0.5em}
\begin{subtable}[t]{0.49\textwidth}
\centering
\caption{Transcription factors - ``embryonic stem cell'' - 2004}
\label{tab:ips}
{\footnotesize
\begin{tabular}{c c c c}
\hline
\trc{Term} & Term + KP & Term  & Co-occur \\
           & Count     & Count & Ratio    \\
\hline

\rowcolor{lg}
*NANOG	& 15	& 59	& 0.254 \\

*UTF1	& 5	& 25	& 0.200 \\

CBX4	& 2	& 21	& 0.095 \\

\rowcolor{lg}
*POU5F1	& 24	& 272	& 0.088 \\

EZH1	& 2	& 25	& 0.080 \\

SOX1	& 8	& 103	& 0.078 \\

IRX4	& 2	& 26	& 0.077 \\

*FOXD3	& 4	& 54	& 0.074 \\

MYF6	& 5	& 79	& 0.063 \\

HOXB4	& 8	& 158	& 0.051 \\

LMO2	& 12	& 240	& 0.050 \\

\rowcolor{lg}
*SOX2 & 11	& 230	& 0.048 \\

EOMES	& 3	& 65	& 0.046 \\

LMX1B	& 5	& 112	& 0.045 \\

LHX2	& 3	& 76	& 0.040 \\

HOXD9	& 3	& 78	& 0.039 \\

HOXD11	& 3	& 80	& 0.038 \\

OTX1	& 5	& 140	& 0.036 \\

HAND1	& 4	& 117	& 0.034 \\

HOXB3	& 3	& 88	& 0.034 \\
\hline
\end{tabular}
}
\end{subtable}
\hfill
\begin{subtable}[t]{0.49\textwidth}
\centering
\caption{Transcription factors - ``cardiomyocyte'' - 2008}
\label{tab:cardiomyocyte}
{\footnotesize
\begin{tabular}{c c c c}
\hline
\trc{Term} & Term + KP & Term  & Co-occur \\
           & Count     & Count & Ratio    \\
\hline
MESP1	& 26	& 89	& 0.292 \\

THRAP1	& 4	& 15	& 0.267 \\

*TBX20	& 30	& 114	& 0.263 \\

\rowcolor{lg}
*GATA4	& 302	& 1294	& 0.233 \\

\rowcolor{lg}
*NKX2-5	& 122	& 528	& 0.231 \\

\rowcolor{lg}
*TBX5	& 104	& 481	& 0.216 \\

GATA5	& 40	& 194	& 0.206 \\

\rowcolor{lg}
*MEF2C	& 151	& 825	& 0.183 \\

\rowcolor{lg}
*HAND2	& 52	& 297	& 0.175 \\

CSRP3	& 8	& 46	& 0.174 \\

IRX4	& 10	& 64	& 0.156 \\

HDAC9	& 26	& 179	& 0.145 \\

NFATC4	& 23	& 173	& 0.133 \\

*IRX5	& 8	& 68	& 0.118 \\

MKL2	& 5	& 43	& 0.116 \\

ISL1	& 51	& 470	& 0.109 \\

*GATA6	& 55	& 526	& 0.105 \\

*HAND1	& 30	& 292	& 0.103 \\

HES2	& 6	& 60	& 0.100 \\

TBX18	& 7	& 73	& 0.096 \\
\hline
\end{tabular}
}
\end{subtable}
\hfill
\vspace{1.5em}
\begin{subtable}[t]{0.49\textwidth}
\centering
\caption{Transcription factors - ``hepatocyte'' - 2009}
\label{tab:hepatocyte}
{\footnotesize
\begin{tabular}{c c c c}
\hline
\trc{Term} & Term + KP & Term  & Co-occur \\
           & Count     & Count & Ratio    \\
\hline
\rowcolor{lg}
HNF1A	& 781	& 849	& 0.920 \\

HNF1B	& 639	& 699	& 0.914 \\

\rowcolor{lg}
*HNF4A	& 466	& 596	& 0.782 \\

*ONECUT1	& 105	& 140	& 0.750 \\

HNF4G	& 23	& 36	& 0.639 \\

\rowcolor{lg}
*FOXA3	& 137	& 217	& 0.631 \\

ONECUT3	& 6	& 10	& 0.600 \\

*FOXA1	& 313	& 571	& 0.548 \\

\rowcolor{lg}
*FOXA2	& 523	& 1055	& 0.496 \\

TCF2	& 136	& 276	& 0.493 \\

MLX	& 325	& 687	& 0.473 \\

*NR0B2	& 54	& 138	& 0.391 \\

*NR1I3	& 66	& 171	& 0.386 \\

*NR1H4	& 66	& 171	& 0.386 \\

HMBOX1	& 5	& 13	& 0.385 \\

NR1I2	& 74	& 200	& 0.370 \\

ONECUT2	& 14	& 40	& 0.350 \\

TCF1	& 137	& 460	& 0.298 \\

*CREB3L3	& 7	& 25	& 0.280 \\

CUTL2	& 13	& 47	& 0.277	\\
\hline
\end{tabular}
}
\end{subtable}

\end{table}

Table \ref{tab:ips} shows the top 20 ranked transcription factors from our query using a list of 2,243 transcription factors and the key phrase ``embryonic stem cell,'' censored to a publication date range through 2004.
Factors that match the landmark papers for producing iPS cells are highlighted gray, and factors that match Mogrify's list of predicted factors are marked with *.
Note that our naive approach is able to reproduce a sufficient list of factors (NANOG, POU5F1, and SOX2) for direct reprogramming\cite{ips2008} in the top 12 hits.
Additionally, five of the top 20 match the list of 70 factors produced by Mogrify.

Table \ref{tab:cardiomyocyte} shows the top 20 ranked hits from our query using a list of transcription factors and the key phrase ``cardiomyocyte,'' censored to a publication date range through 2008.
Again, factors that match the landmark papers for direct reprogramming to cardiomyocytes are highlighted in gray, and factors that match Mogrify's list of predicted factors are marked with *.
Similar to the iPS cell query, our approach reproduces the complete list of early published transcription factors in the first nine hits, and nine of the top 20 hits match the list of 57 factors predicted by Mogrify.

\begin{table}[!ht]
\centering
\caption{Top 50 drug and device hits for our drug repositioning query against the key phrase ``hypoglycemia.''  Hits that are diabetes drugs are not highlighted.  Hits that are not diabetes drugs, but which are known to decrease blood sugar are highlighted in green, and hits that increase blood sugar are highlighted in red.  Hits that are not diabetes drugs, but were also not in our labeled list, are highlighted in gray.  Hits that are exact matches to those in Kuang et al.\cite{cdr2016charles} are marked with *.}
\label{tab:drugs}
\footnotesize
\begin{tabular}{c c c c}
\hline
\trc{Drug} & Drug + KP & Drug  & Co-occur \\
           & Count     & Count & Ratio    \\
\hline
GLYBURIDE MICRONIZED	& 3	& 4	& 0.750 \\

GLYNASE	& 16	& 27	& 0.593 \\

MICRONASE	& 24	& 41	& 0.585 \\

NOVOLIN N	& 28	& 48	& 0.583 \\

STARLIX	& 26	& 46	& 0.565 \\

TOLINASE	& 14	& 26	& 0.538 \\

GLIPIZIDE XL	& 7	& 13	& 0.538 \\

*GLUCOTROL XL	& 15	& 28	& 0.536 \\

*INSULIN DETEMIR	& 547	& 1107	& 0.494 \\

PREMEAL	& 477	& 975	& 0.489 \\

SUBCUTANEOUS INSULIN INFUSION PUMP	& 37	& 76	& 0.487 \\

*INSULIN ASPART	& 723	& 1509	& 0.479 \\

*INSULIN LISPRO	& 717	& 1515	& 0.473 \\

NPH INSULIN	& 787	& 1665	& 0.473 \\

PRECOSE	& 31	& 66	& 0.470 \\

PRANDIN	& 27	& 59	& 0.458 \\

LANTUS	& 290	& 640	& 0.453 \\

*GLUCOTROL	& 41	& 91	& 0.451 \\

NOVOLOG	& 90	& 203	& 0.443 \\

\rowcolor{up}
ZESTORETIC	& 3	& 7	& 0.429 \\

*HUMALOG	& 210	& 495	& 0.424 \\

*AMARYL	& 47	& 113	& 0.416 \\

INSULIN NPH	& 281	& 691	& 0.407 \\

GLYBURIDE-METFORMIN	& 46	& 117	& 0.393 \\

REGULAR INSULIN	& 1182	& 3048	& 0.388 \\

*GLIMEPIRIDE	& 935	& 2487	& 0.376 \\

BYETTA	& 136	& 370	& 0.368 \\

\rowcolor{down}
ZEBETA	& 4	& 11	& 0.364 \\

HUMULIN N	& 50	& 140	& 0.357 \\

*GLUCOPHAGE XR	& 16	& 45	& 0.356 \\

PRAMLINTIDE ACETATE	& 21	& 60	& 0.35 \\

JANUVIA	& 87	& 252	& 0.345 \\

LIRAGLUTIDE	& 885	& 2589	& 0.342 \\

INSULIN REGULAR HUMAN	& 41	& 121	& 0.339 \\

\rowcolor{up}
AVALIDE	& 3	& 9	& 0.333 \\

\rowcolor{up}
DEMADEX	& 3	& 9	& 0.333 \\

NATEGLINIDE	& 351	& 1098	& 0.320 \\

REPAGLINIDE	& 458	& 1486	& 0.308 \\

AVANDAMET	& 17	& 56	& 0.304 \\

*EXENATIDE	& 1136	& 3843	& 0.296 \\

*GLIPIZIDE	& 669	& 2278	& 0.294 \\

GLUCAGEN	& 29	& 99	& 0.293 \\

*BLOOD-GLUCOSE METER	& 539	& 1926	& 0.280 \\

WELCHOL	& 22	& 79	& 0.278 \\

\rowcolor{down}
TIAZAC	& 4	& 15	& 0.267 \\

GLUCOVANCE	& 18	& 68	& 0.265 \\

\rowcolor{lg}
TEQUIN	& 11	& 43	& 0.256 \\

NOVOLIN R	& 68	& 270	& 0.252 \\

NOVOLIN	& 143	& 570	& 0.251 \\

\rowcolor{down}
CALAN SR	& 4	& 16	& 0.25 \\
\hline
\end{tabular}
\end{table}

Table \ref{tab:hepatocyte} shows the top 20 ranked hits from our query using a list of transcription factors and the key phrase ``hepatocyte,'' censored to a publication date range through 2009.
Again, factors that match the landmark papers for direct reprogramming to hepatocytes are highlighted in gray, and factors that match Mogrify's list of predicted factors are marked with *.
KinderMiner successfully reproduces four of the six factors for maturation from the landmark literature, and nine of the top 20 hits match the 27 predicted by Mogrify.

Table \ref{tab:drugs} shows the top 50 drugs and devices ranked by our method as being relevant to hypoglycemia (low blood sugar).
Drugs that are advertised specifically to treat diabetes are not highlighted.
The highlighted drugs are not specifically advertised to treat diabetes.
Drugs highlighted green are labeled as drugs that may reduce blood sugar, drugs highlighted red may increase blood sugar, and drugs highlighted gray are not present in our labeled list.\cite{diabetesincontrol}

Perhaps unsurprisingly, 43 of our top 50 hits are specifically for treatment of diabetes, due in part to the abundance of diabetes drugs and various brand names thereof.
We note that the hit \emph{premeal} is likely a result of correlation to premeal insulin.
These 43 hits are a positive result as they suggest that our method successfully finds relevant correlations, but the more interesting hits are those that are not diabetes drugs as our goal is to reposition drugs.
Of the seven hits that are not specifically diabetes drugs, Zestoretic, Avalide, and Demadex have been shown to potentially increase blood glucose, whereas Zebeta, Tiazac, and Calan SR have been shown to potentially decrease blood glucose.
Tequin is not in our labeled list.
It is an antibiotic that has been shown to increase patient risk of dysglycemia (either hypoglycemia or hyperglycemia).\cite{tequin2006}

Overall, in all of our evaluation tasks, our method finds numerous relevant hits and demonstrates overlap with the results of far more sophisticated methods designed specifically for the separate tasks presented.

\section*{Conclusions and Future Work} \vspace{-1em}
In this work, we present a simple and general text mining method for predicting pairwise associations between a key phrase and target terms.
We demonstrate the use of this method for identifying transcription factors that are important for three cell reprogramming tasks and for discovering candidate drugs for alternative uses.
In both of our application domains, we find that KinderMiner identifies numerous relevant hits and overlaps with state-of-the-art methods designed specifically for each domain.
In historically censored searches of factors for reprogramming cell states, KinderMiner highly ranks transcription factors that would, years later, be shown to be important for reprogramming to cell states of interest, thus providing a short, ordered list of candidates for biologists that would have greatly simplified the challenging combinatorial task they faced. 
Importantly, the domain-specific approaches require domain-specific data, whereas KinderMiner only requires an indexed text corpus.
We argue that our method is a valuable new tool that can be used to help prioritize research directions despite its naivet\'{e}.
Furthermore, we anticipate that our method may prove valuable in domains other than biomedicine by mining other large plain text corpora.

We view the simplicity of KinderMiner as a strength, but this simplicity also leads to limitations.
For example, KinderMiner does not explicitly implement any actual natural language processing.
Thus, terms like the transcription factor T (Brachyury) are likely to match many articles that do not reference the T gene, but may in fact be matches to middle initials or similar.
We do not observe this particular phenomenon in our lists of top 20 hits presented here, but we anticipate that this may be a problem for other queries.
While we believe there is value in the simplicity of our method, we expect that the addition of techniques such as text normalization and named entity recognition may help alleviate this issue and, therefore, propose it as future work.

Furthermore, we observe that some of our queries have low total counts of articles for sorting by ratio.
For example, THRAP1 in Table \ref{tab:cardiomyocyte} counts a total of 15 articles that contain the term, four of which contain both the term and key phrase.
This may pose a greater challenge when using smaller corpora, or when querying terms or key phrases that are relatively new within the literature.
A query that counts a total of four articles, three of which have both term and key phrase may be ranked well by ratio, but is unlikely to actually represent compelling evidence of association.
In general, there will always be a horizon of discovery defined by the quantity of published literature for particular key phrases and target terms, but we will explore the use of thresholding, pseudocounts, and other Bayesian approaches to modulate the rank of such cases in future work.

Finally, we note that constructing a search engine around large corpora is non-trivial.
We were fortunate with our applications in that Europe PMC offers a web API on which we built KinderMiner, but not all corpora will afford such convenience.
We do not propose any specific suggestions for how to address this issue, but instead expect that time will assist with the continued democratization of search tools (e.g. Apache Lucene and SOLR).
We anticipate that the availability of easy-to-use software packages will continue to grow, and we propose evaluating applications of KinderMiner using such software on open data as future work.

\section*{Acknowledgements} \vspace{-1em}
The authors acknowledge support from the National Institutes of Health (NIH) grant number UH3TR000506-05 and the National Institute of General Medical Sciences (NIGMS) grant number R01GM097618-05.  The authors also thank Marv and Mildred Conney for a grant to R.S. and J.A.T., and Erin Syth for editorial assistance.

\bibliographystyle{vancouver}
\center{\textbf{References}}
{\small
{\def\section*#1{} 
{\linespread{0.5}\selectfont
\bibliography{amiacri17_kindermining_paper}
\par} 
}
} 

\end{document}